**Anomalous doping evolution of nodal dispersion revealed by *in-situ* ARPES on continuously doped cuprates**

Yigui Zhong[1,2,*], Jianyu Guan[1,2,*], Jin Zhao[1,2], Cenyao Tang[1,2], Zhicheng Rao[1,2], Haijiang Liu[1,2], Jianhao Zhang[3], Sen Li[3], Zhengyu Weng[3], Genda Gu[4], Yujie Sun[1,5,†], Hong Ding[1,2,5,†]

[1]*Beijing National Laboratory of Condensed Matter Physics and Institute of Physics, Chinese Academy of Sciences, Beijing 100190, China*
[2]*School of Physics, University of Chinese Academy of Sciences, Beijing 100049, China*
[3]*Institute for Advanced Study, Tsinghua University, Beijing 100084, China*
[4]*Condensed Matter Physics and Materials Science Department, Brookhaven National Laboratory, Upton, New York 11973, USA*
[5]*Songshan Lake Materials Laboratory, Dongguan, Guangdong 523808, China*

[*]*Yigui Zhong and Jianyu Guan contributed equally to this work.*

[†] *Corresponding authors: dingh@iphy.ac.cn; yjsun@aphy.iphy.ac.cn*

Abstract

We study the systematic doping evolution of nodal dispersions by *in-situ* angle-resolved photoemission spectroscopy on the continuously doped surface of a high-temperature superconductor $Bi_2Sr_2CaCu_2O_{8+x}$ and reveal that the nodal dispersion has three fundamentally different segments separated by two kinks, located at ~10 meV and roughly 70 meV, respectively. These three segments have different band velocities and different doping dependence. In particular, in the underdoped region the velocity of the high-energy segment increases monotonically as the doping level decreases and can even surpass the bare band velocity. We propose that electron fractionalization is a possible cause for this anomalous nodal dispersion and may even play a key role in the understanding of exotic properties of cuprates.

For the high-temperature (high-$T_c$) cuprates, one of the puzzling features is the "kink" in the electronic band dispersion, which universally exists in different superconducting families of cuprates [1-3]. Along the nodal direction of the d-wave superconducting gap, the kink appears at a binding energy of roughly 70 meV in many cuprates [1, 4-6]. Meanwhile near the antinode with the maximum energy gap, a seemingly stronger kink is located at about 20-40 meV [7-9], depending on the doped carrier concentration $x$. Many scenarios have been proposed to understand the origin of this kink feature [10-12]. For examples, one popular proposal is that a strong electron-phonon coupling induces a kink in the dispersion [13, 14], while another one suggests that the electron-magnetic mode coupling is responsible for the kink [15-17]. Yet, the origin of this universal kink is far from settled. Recently we developed a new technique to continuously change the doping level of surface layers by annealing sample in ozone or vacuum atmosphere, which enables a systematic *in-situ* angle-resolved photoemission spectroscopy (ARPES) study on the same surface with a wide range of doping level, thus promising more precise studies on doping evolution of many important properties of the cuprates [18].

In this letter, we systematically study the doping evolution of the nodal dispersion by performing *in-situ* ARPES measurements on a continuously doped surface of $Bi_2Sr_2CaCu_2O_{8+x}$ (Bi2212). From the measured band dispersions, two kinks are identified at ~10 meV and 45-70 meV below the Fermi level ($E_F$), respectively, and they separate the nodal dispersion into three segments with different band velocities. We clearly demonstrate the fundamentally different doping evolutions for these three band velocities: a decreasing velocity for the lower segment, a no-change velocity for the middle segment, and an increasing velocity for the higher segment, as the doping level is decreased. What is more exotic is that the velocity of the higher segment becomes divergent as doping decreases and can even exceed the bare band velocity for the extremely underdoped (UD) ones. After discussing several possible scenarios proposed to explain the kink phenomena, we suggest that electron fractionalization at the high energy is likely responsible for the anomalous doping evolution of the nodal dispersion in the UD region.

Sample preparation and surface treatment of ozone/vacuum annealing to continuously change the doping level of surface layers were described in our previous paper [18]. *In-situ* ARPES measurements were performed in a laboratory ARPES system equipped with a Scienta R4000 analyzer and a Scienta VUV light source. He-Iα resonant line (hν = 21.218 eV) was used and the vacuum of the ARPES chamber was better than $3 \times 10^{-11}$ Torr. The energy and angle resolution were set as ~5 meV and ~0.2°, respectively.

We first show the spectrum image plots with the same color scale along the nodal direction at various different doping levels from $x \sim 0.24$ to 0.08 [Figs. 1(a) – 1(f)], which were acquired on the same surface through annealing sample under the ozone/vacuum circumstance. The well-known 70-meV kink is clearly observed at each doping level, and the spectral intensity has a sudden reduction after crossing the kink. The area with high intensity (light color in images) shrinks dramatically when it goes to the UD region. Interestingly, the band beyond the kink becomes more and more vertical, which indicates that the kink becomes stronger with decreasing doping level. In order to extract the band dispersion for quantitative comparison, a standard practice [19] is to perform a Lorentzian fitting to momentum distribution curves (MDCs) that are intensity distributions as a function of momentum at a fixed energy [Figs. 1(g) – 1(i)], with an assumption that the self-energy in Green's function has no or weak dependence on momentum. From the fitting, the peak positions of MDCs trace out the band dispersion and the widths contain the information of the quasiparticle lifetime. The extracted band dispersion for each doping level is plotted with the black lines in Figs. 1(a)-1(f). Again, one can clearly see that the kink phenomena become stronger with decreasing doping level. The spectral width is also seen to become wider as the doping level decreases [Figs. 1(g)-1(i)]. Meanwhile, the spectral width becomes larger suddenly when it passes across the kink [Figs. 1(j)-1(k)], indicating an abrupt change of the quasiparticle lifetime.

In principle, ARPES probes the single-particle spectral function, which treats the many-body interactions as additional self-energy renormalization to electrons [20]. One way to understand the interactions in a material is to analyze the self-energy $\Sigma(\omega)$ from ARPES spectra. By fitting the measured band and comparing with its corresponding bare band [Fig. 2(a)], one can extract the real part of self-energy Re$\Sigma$, which is the energy difference between the renormalized band and the bare band, and the imaginary part of self-energy Im$\Sigma$, which is proportional to the MDC width. According to previous practices [21], the low-energy bare band along the nodal direction in the cuprates can be approximated as a linear line that connects the measured Fermi crossing point and the higher binding energy, typically around ~200 meV. We note that there are other different forms for the bare band used in previous literature [22]. For comparison, the MDC-derived nodal dispersions at different doping levels are plotted with the relative momentum to their corresponding Fermi momenta ($k_F$) [Figs. 2(a) and 3(a)]. We notice that previous literature [1, 4-9] usually plot the Re$\Sigma$ and the MDC width in the energy coordinate due to the weak dependence on momentum. Here for the better visualization we plotted them against the peak positions of their corresponding MDCs [Figs. 2(b) and 2(c)]. Like the regular plots in the energy coordinate, one can see the conspicuous peaks in Re$\Sigma$ curves [Fig. 2(b)], which indicate that there is a large renormalization of the electronic dispersion at the place where the kink happens. The relative momentum position of the kink, $\Delta k(x) = k_{kink} - k_F$, can be identified through this conspicuous peak. With decreasing doping level, this peak in Re$\Sigma$ curves gradually grows and moves to the $k_F$ and its height is enhanced due to the stronger kink. Meanwhile the width of MDCs dramatically increases after the kink for the UD samples, while it is much milder before the kink [Fig. 2(c)]. It resembles that electrons are undergoing a decay from a well-defined quasiparticle to something with an extremely shorter lifetime when they go cross the kink. This behavior becomes more and more noteworthy as the doping level decreases, especially in the UD side [Fig. 2(c)]. What is the more impressive is that the kink gradually moves toward their corresponding $k_F$ and $E_F$ with decreasing doping level in the UD side, which directly suggests the area with a well-defined quasiparticle shrinks [Figs 2(d) and 2(e)].

Since the quasiparticle lifetime experiences a large transformation after crossing the kink, we next study the effective mass of the band dispersion before and after the kink. The MDC-derived nodal band dispersions at different doping levels are summarized in Figs. 3(a)-3(c). Besides the kink at 45-70 meV, one can see another kink of the band dispersion at a lower binding energy ~10 meV, which contributes a "shoulder" in the ReΣ curves [Fig. 2(b)] and becomes more pronounced for the UD samples [Fig. 3(a)]. These two kinks separate the nodal band dispersion into three segments with different band velocities or effective masses. To make the results more precise, we avoid the vicinity of these two kinks and evaluate velocities of three band segments at the binding energy of 0-5 meV, 25-55 meV and 80-125 meV [Fig. 3(d)]. The numerical analysis of doping dependence on the velocities of three band segments is displayed in Fig. 3(e). We find that the velocity of the segment just below $E_F$ ($S_L$, 0-5 meV) decreases smoothly with decreasing doping level [Fig. 3(b)], which is consistent with the theoretical expectation that stronger electron-electron interactions at UD samples shall make electrons heavier [23]. The velocity of the middle segment ($S_M$, 25-55 meV) remains a constant within error-bar [Fig. 3(c)] over a wide doping range, consistent with previous observations [24, 25]. The exotic behavior is that the velocity of the segment at the higher binding energy ($S_H$, 80-125 meV) dramatically increases with decreasing doping level in the UD region [Fig. 3(a)], in opposite with the expected trend. More surprisingly, at the heavily UD level the velocity of the higher binding energy segment can even surpass its bare band velocity which is defined by LDA calculations [26, 27]. Those exotic behaviors of the velocity imply an unconventional origin of the kink in the nodal dispersion.

The popular scenario of kink origin attributes these phenomena in electronic dispersion as the renormalized effect to the electrons due to the electrons coupling with some kinds of bosonic modes at specific energies, like phonons or spin resonance [1, 13-17]. Although some of these bosonic modes do exist in the cuprates [28-30], these coupling scenarios are unable to explain the doping dependence of the kink phenomena, especially in the aspect of the divergent band velocities of $S_H$. The magnetic resonance mode scenario is unlikely since the kink in the nodal dispersion still robustly survives in the normal states where the magnetic resonance mode vanishes [1, 30], so here we focus on discussing the electron-phonon coupling scenario. Firstly, under the frame of the electron-phonon coupling, typically the bare bands can be regarded approximately as a linear line within a narrow energy window that connects the measured $E_F$ and a higher binding energy just like the dashed line in Fig. 2(a). Thus, with decreasing doping level, due to the increasing velocities of $S_H$, the bare bands will become more and more vertical, which is directly incompatible with the simple rigid band shift picture of Bi2212 [31, 32]. Secondly, according to the weak electron-phonon coupling scenario described by Migdal-Eliashberg picture [33], since the electron-phonon coupling domain should be restricted to a small energy range around the typical energy of phonon modes ($\Omega$), so the dispersion well above $\Omega$ tends to be recovered to the position of the noninteracting band. Thus the band velocity of the segment at a higher binding energy than $\Omega$ will remain relatively stable, and for the binding energy lower than $\Omega$, the corresponding velocity will change with the coupling strength, which is opposite to what was observed here. Thirdly, we noticed there is a the strong electron-phonon coupling limit based on the polaron picture [34, 35] which is argued to be a proper scenario for undoped and lightly underdoped cuprates. It argued that the Fermi energy in an undoped or lightly underdoped sample is relatively small compared with the energy of optical phonons, therefore the multi-phonon processes can, in principle, lead to the small polaron formation, thus they attributed the kink dispersion to the scattering between doped holes and small polarons [35]. However, a placid change of quasiparticle lifetime at the kink point and a more placid evolution of high-energy band velocity as a function of doping level inferred from polaronic scenario cannot capture our measurements here.

On the contrary, it is reasonable that the kink at the lower energy ~10 meV stems from the electron-phonon coupling [25, 36, 37] due to that it fits with the conventional coupling frame. The $S_M$ after the lower energy kink recovers to the non-coupling dispersion which has a constant velocity, while distinguishably the $S_L$ before the lower energy kink have different velocities due to variable strength of coupling. In fact, there is strong evidence of the existence of the lower energy phonons in Bi2212 [38].

We noticed that Randeria et al [11] phenomenologically suggested the velocities of $S_H$ are a consequence of the renormalization to the nodal quasiparticle weight $Z_k$ through a formula $V_{high} = V_{low}/Z_k$ , where $V_{low}$, which is constant, is the velocity of $S_M$. According to this suggestion, the nodal quasiparticle weight $Z_k$ should decrease quite rapidly with decreasing doping level, which is not consistent with our previous results that the nodal quasiparticle weight $Z_k$ keeps relatively constant over a wide doping range [18].

The unconventional behavior of the velocity of $S_H$ surpassing the predicted velocity of the bare band is reminiscent of the electron fractionalization in some of the 1D materials [39, 40]. Similar scenarios of electron fractionalization have also been proposed for the 2D cuprates [41, 42]. For the parent compound of a cuprate superconductor, a Mott insulator, the charge is gapped due to strong on-site repulsive energy [43], while the spin has a gap-less excitation [44], so the charge and spin degrees can be separated in the low-energy excited processes, namely, the electrons are fractionalized. With the holes doped into, the system gradually evolves into a high-$T_c$ superconductor, and this fractionalization effect becomes weaker but still leaves some signatures in the UD region. In fact, there are many theoretical models suggesting that the spin-charge separation is the key to the high-$T_c$ cuprates [43, 45-47]. One of these theoretical models is the so-called phase string theory [47]. In a doped Mott insulator described by the simplest t-J model [43], the doped holes cannot propagate coherently due to the presence of a nontrivial sign structure stemming from the phase string sign structure, which is distinct from the conventional fermion sign structure [47-49]. In order to incorporate the charge propagation properly, the doped holes should be fractionalized at a higher energy. The phase string theory [50] predicts that a peculiar electron fractionalization happens in the doping region accompany with the pseudogap, which intrinsically leads to a two-component structure: 1. The band segment of lower energy before the 70 meV kink (especially $S_M$) would be the "protected" emergent quasiparticle with the bare-band Fermi velocity, which is independent with the doped concentration $x$; 2. The higher segment $S_H$ after the kink acquires a larger band velocity due to the natural electronic fractionalization, which increases conspicuously with decreasing doping level, especially for very low doped ones. Naturally, due to the electronic fractionalization, the MDC width has a sudden increase after crossing the kink. A comprehensive theoretical demonstration along this line is given in a separated paper [51], which nicely reproduced many observed features in our ARPES measurements.

Actually, the distinct behaviors of the quasiparticle before and after the kink expose the happening of electron fractionalization at the high energy. Hereby we comparing the EDCs before the kink (extracted at $k - k_f = 0.017 \text{ Å}^{-1}$) and after the kink (extracted at $k - k_f = 0.043 \text{ Å}^{-1}$) [Fig. 1(k)], one can clearly see that for the EDCs before the kink, due to the "protected" quasiparticles as the described above, the quasiparticle peaks can be clearly recognized at all doping levels, however, for the EDCs after the kink, the broaden peaks due to the highly electronic scattering are gradually melted into the incoherent background when it goes to the UD side, simultaneously the band velocities at the high energy are extremely diverged. On the other hand, the peaks at MDCs are persistently recognized no matter before or after the kink, while dramatically become broadening after the kink with the doping decreases [Fig. 1(j)]. As suggested by Laughling [52] and Orgad et al [53], the no-peak feature at EDCs but still peak feature in MDCs can be viewed as a spectral evidence of electron fractionalization. For our case, after excluding the influences from surface disorder, resolution effect, and background signal through our *in-situ* surface doping technique [18], hereby we argue that our observation, the vanishing behavior of EDCs' peaks with robust peak-feature in MDCs [Figs. 1(j) and (k)] after the kink, indicates a possible electron fractionalization happening at the high energy, which causes the kink phenomena at the dispersions.

In conclusion, we conduct a systematic ARPES study of nodal dispersion of Bi2212 over a wide doping range, and reveal that the velocity of the segment after the high-energy kink unconventionally increases with decreasing doping level in the UD region, which cannot be explained by the electron-mode coupling scenario. Alternatively, we propose the electronic fractionalization effect at the high energy might be responsible for the unconventional behaviors.

Figures

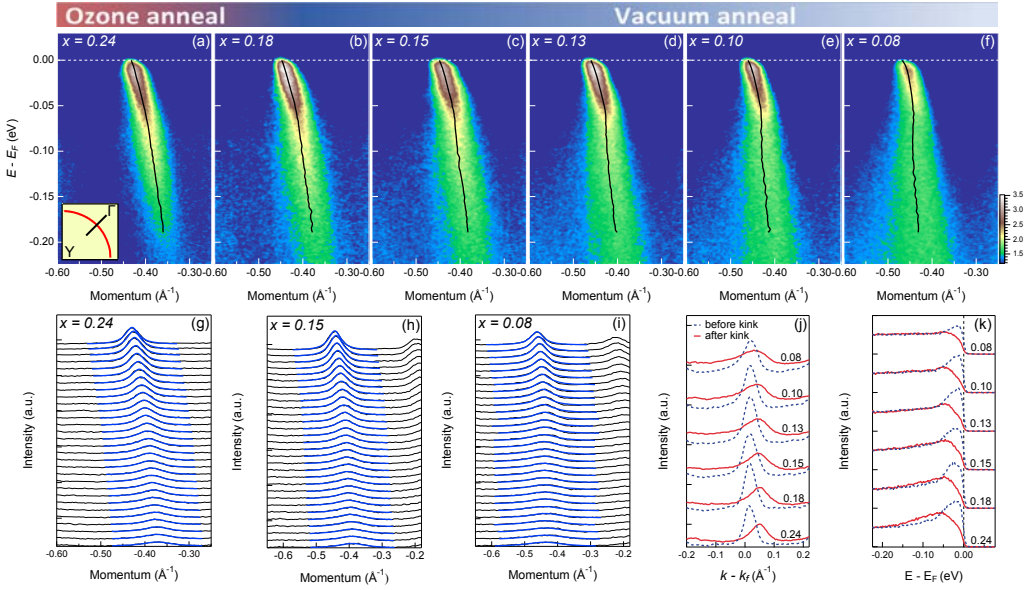

FIG. 1. (Color online) (a) ARPES spectrum of ozone-annealed Bi2212 along the nodal direction (as illustrated in the inset), the doping level is estimated as 0.24. Here ozone anneal means sample is annealed at ozone atmosphere. (b) - (f) Spectra acquired from the same sample surface after annealed in vacuum with different temperatures (called vacuum anneal) step by step, and the corresponding doping levels are estimated as 0.18, 0.15, 0.13, 0.10, and 0.08, respectively. These spectra of (a) - (f) are plotted in the same color scale which is illustrated in the lower right side of (f). (g) - (i) The black lines are the MDC plots for the doping levels at 0.18, 0.15 and 0.10, respectively, and the blue lines are their Lorentzian fittings. (j) The MDCs before the kink (blue dash lines, $E - E_F = -27\ meV$) and after the kink (red solid lines, $E - E_F = -105\ meV$) at different doping levels. (k) The EDCs before the kink (blue dash lines, $k - k_f = 0.017\ Å^{-1}$) and after kink (red solid line, $k - k_f = 0.043\ Å^{-1}$) at different doping levels.

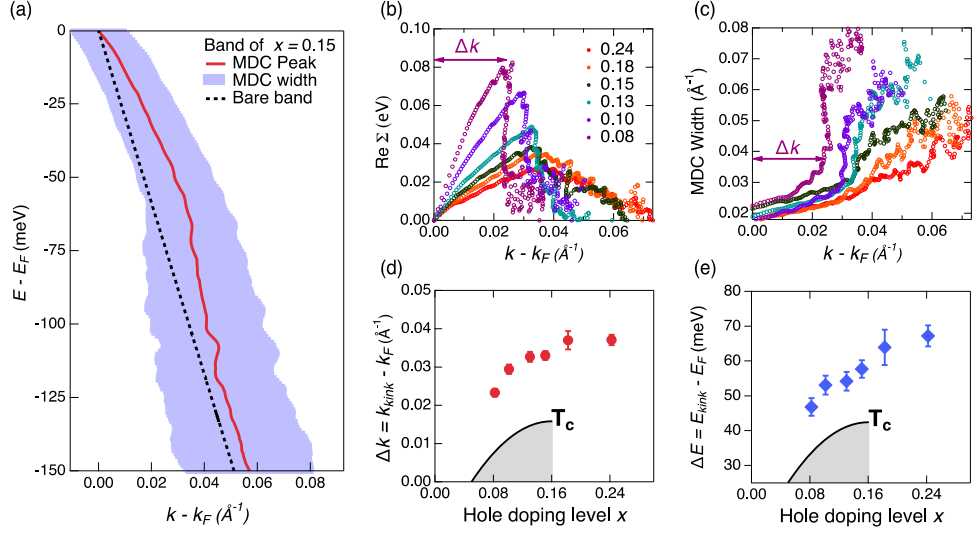

FIG. 2. (Color online) (a) MDC-derived band dispersion at the doping level equals 0.15: the red line is the peak of MDCs extracted by Lorentzian fitting, and the violet shadow represents the width of each MDC. The dashed line is a typical bare band as described in the text. (b) The real parts of self-energy at each doping level, which are plotted against with the corresponding peaks of each MDC recorded using the relative momentum $k - k_F$. (c) The MDC widths of the spectrum at each doping level, which are also plotted against with the corresponding MDC peaks recorded using the relative momentum $k - k_F$. (d) - (e) The determined momentum positions $\Delta k(x)$ and energy positions $\Delta E(x)$ for each doping level, which are relative to their corresponding Fermi momenta or Fermi level.

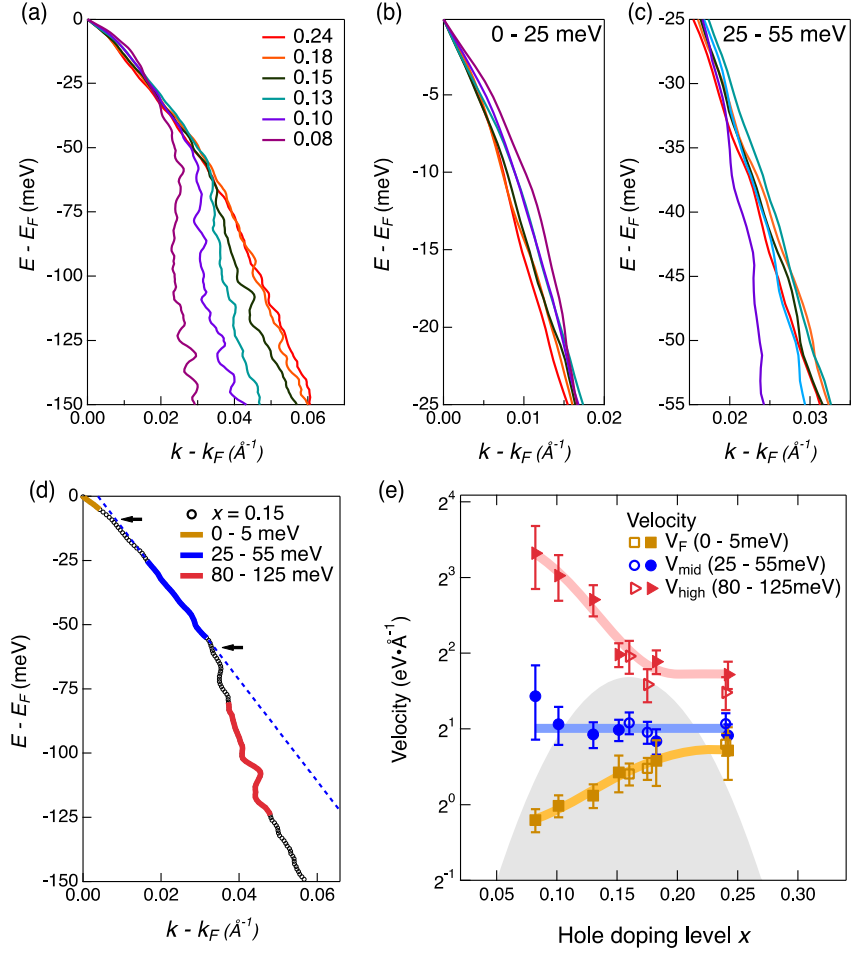

FIG. 3. (Color online) (a) Summary of MDC-derived nodal dispersions at different doping levels from 0.24 to 0.08 referred in this paper. The momentum is scale to their corresponding Fermi momentum. (b) The MDC-derived nodal dispersion segments at 0-25 meV below $E_F$. (c) The MDC-derived nodal dispersion segments at 25-55 meV. (d) Three segments separated by two kinks which are marked with the black arrows: the brown one is at the 0-5meV, the blue one is at the 25-55 meV, the red one is at the 80-125 meV. The black circles are the nodal dispersion at the doping level equals 0.15. (e) The doping evolution of velocities of three segments plotted with the corresponding colors marked in (d). The empty markers with corresponding shapes represent an independent dataset acquired from another sample. The colored shadow lines are guides for eyes. The gray shadow dome represents the superconducting phase region.

**Acknowledgment**

We thank J.-P Hu and Z.-Q. Wang for helpful discussion and thanks F. -Z. Yang and J. -R. Huang for technique assistant. The work at the Institute of Physics is supported by the grants from the Ministry of Science & Technology of China (2016YFA0401000, 2015CB921300, 2015CB921000), the Natural Science Foundation of China (11227903, 11574371), the Chinese Academy of Sciences (XDB07000000, XDPB08-1), and Beijing Municipal Science & Technology Commission (Z181100004218005). Genda Gu is supported by the Office of Basic Energy Sciences, Division of Materials Science and Engineering, U.S. Department of Energy, under Contract No. DE-SC0012704.